\documentstyle[12pt,epsf,epsfig]{article}
\def\be{\begin{equation}}
\def\ee{\end{equation}}
\def\ba{\begin{eqnarray}}
\def\ea{\end{eqnarray}}
\def\la{~\mbox{\raisebox{-.6ex}{$\stackrel{<}{\sim}$}}~}

\def\bq{\begin{quote}}
\def\eq{\end{quote}}

 at 10truept

\newcommand{\beq}{\begin{equation}}
\newcommand{\eeq}{\end{equation}}
\newcommand{\beqa}{\begin{eqnarray}}
\newcommand{\eeqa}{\end{eqnarray}}

\def\la{~\mbox{\raisebox{-.6ex}{$\stackrel{<}{\sim}$}}~}

\def\ltap{\ \raise.3ex\hbox{$<$\kern-.75em\lower1ex\hbox{$\sim$}}\ }
\def\gtap{\ \raise.3ex\hbox{$>$\kern-.75em\lower1ex\hbox{$\sim$}}\ }
\def\gl{\ \raise.5ex\hbox{$>$}\kern-.8em\lower.5ex\hbox{$<$}\ }
\def\roughly#1{\raise.3ex\hbox{$#1$\kern-.75em\lower1ex\hbox{$\sim$}}}

\begin{document}

\thispagestyle{empty}
\begin{flushright}
astro-ph/0406099\\ June 2004
\end{flushright}
\vspace*{1cm}
\begin{center}
{\Large \bf Observational Implications of Cosmological}\\
\vskip.2cm
{\Large \bf Event Horizons }\\
\vspace*{1.5cm} {\large Nemanja Kaloper$^{\dagger,}$\footnote{\tt
kaloper@physics.ucdavis.edu}, Matthew
Kleban$^{\ddagger,}$\footnote{\tt matthew@ias.edu}
and Lorenzo Sorbo$^{\dagger,}$\footnote{\tt sorbo@physics.ucdavis.edu} }\\
\vspace{.5cm}
{\em $^{\dagger}$Department of Physics, University of California, Davis,
CA 95616}\\
\vspace{.15cm}
{\em $^{\ddagger}$Institute for Advanced Study, Princeton, NJ 08540}\\
\vspace{.15cm} \vspace{2cm} ABSTRACT
\end{center}
In a universe dominated by a small cosmological constant or by
eternal dark energy with equation of state $w < -1/3$, observers
are surrounded by event horizons. The horizons limit how much of
the universe the observers can ever access. We argue that this
implies a bound $N \sim 60$ on the number of e-folds of inflation
that will ever be observable in our universe if the scale of the
dark energy today is $\sim (10^{-3} eV)^4$. This bound is
independent of how long inflation lasted, or for how long we
continue to observe the sky. The bound arises because the imprints
of the inflationary perturbations thermalize during the late
acceleration of the universe. They ``inflate away" just like the
initial inhomogeneities during ordinary inflation. Thus the
current CMB data may be looking as far back in the history of the
universe as will ever be possible, making our era a most opportune
time to study cosmology.

\vfill
\setcounter{page}{0}
\setcounter{footnote}{0}
\newpage

\section{Introduction}

Cosmological observations suggest that the expansion of the
universe may be accelerating \cite{sne,new,cmb}. This may indicate
that the universe contains a dark energy component with equation
of state $w = p/\rho \la -2/3$, comprising as much as $70\%$ of
the critical energy density, $\rho_c \sim (10^{-3} eV)^4$. The
``usual suspects" for dark energy are either a cosmological
constant or a time-dependent quintessence field \cite{weinberg,q}.
If the dark energy includes an eternal component obeying $w <
-1/3$, the universe will continue to accelerate indefinitely. Any
observer in it is surrounded by an event horizon. This limits the
region of the universe accessible to observation, and raises
interesting conceptual problems \cite{tom,hks,witten,cosconst}.

Recently, the authors of \cite{tomwilly} have argued that there is
an incompatibility between the assumptions of the quantum field
theory description of very long inflation and the bound on the
number of states in the Hilbert space of an asymptotically de
Sitter universe. Their conclusion is that in a universe dominated
at late times by a positive, stable, cosmological constant, there
is an upper limit on the number $N$ of e-folds of inflation that
can be described by a conventional quantum field theory. If
inflation were longer, it would deposit so much matter in the
universe that it would collapse to a big crunch at a time after
reheating. However, long inflation was subsequently analyzed in
\cite{marolflowe}, which found no bound on $N$.

In this note we first briefly re-examine the arguments of
\cite{tomwilly}. Adhering to the covariant entropy bounds
\cite{willylenny,bousso} as did \cite{marolflowe}, we do not find
a bound on how many e-folds the early inflation could have lasted
in a universe which starts to accelerate forever. However, we do
find a bound $N \sim 60$ on the number of e-folds that will ever
be {\em observable}. If our universe accelerates forever, we will
never see past the last $60$ e-folds or so.

Further, we find that a universe which transitions to an eternally
accelerating phase will contain the most information about the
inflationary perturbations at the epoch of transition. Later
observers will be able to observe less and less about the
inflationary phase, because the fluctuations generated during
inflation will cease reentering the horizon, and those that did
reenter will be evicted again. The overall amplitude of the CMB
will redshift, and more significantly the pattern of anisotropies
will freeze in such a way that little new information will become
available. Eventually, the CMB will redshift to a point where it
is permanently contaminated by cosmological Hawking radiation.

Therefore, the current cosmological observations may already be
looking as far back in the early universe as may ever be possible,
making this a most opportune time to study cosmology. This
provides an interesting new twist to the ``Why Now?" problem: Why
is now ($\pm$ few current Hubble times) the best time to observe
the signatures of early universe physics?

\section{Counting E-folds}

We first briefly review the argument of
\cite{tomwilly}\footnote{Here we focus only on the case $\kappa =
p/\rho = 1$ of \cite{tomwilly}, which gave the weakest bound.}.
Imagine that the early universe begins as a flat, inflating
universe with Hubble parameter $H_i$. This is a good approximation
soon after the onset of inflation. Consider one initial Hubble
patch, with a volume $\sim {1}/{(H_i)^3}$. After $N$ e-folds of
inflation, the volume of this inflating patch has increased by a
factor of $e^{3 N}$. Suppose that inflation then ends and the
universe reheats, and the decay of the inflaton produces some
local entropy density $\sigma$. In units where the Planck mass is
set to unity, and assuming that the total entropy $S_r$ generated
at reheating is simply the product of the entropy density and the
volume, one obtains:
\be S_r = \sigma \frac{e^{3 N}}{( H_i)^3} \, . \label{entropy} \ee

If there is a positive cosmological constant $\lambda$, the future
of the universe will be either a big crunch or a de Sitter space
with Hubble parameter $H_0 \sim \sqrt{\lambda}$, depending on the
amount of energy released at the end of inflation. If the universe
is to avoid a big crunch, \cite{tomwilly} argues that the total
entropy generated at the end of inflation must be bounded by the
Gibbons-Hawking entropy of the final de Sitter space, which is
given by the horizon area in Planck units. At sufficiently late
times, this is the same as the horizon area of a single
post-inflationary Hubble patch: $S_{GH} \simeq 1/(H_0)^2$.
Requiring $S_r \la {1}/{(H_0)^2}$ yields
\be N \la {2 \over 3} \ln {H_i \over H_0} \sim 85 \, ,
\label{bound} \ee
when  $\sigma \sim H_i$, $\kappa = 1$, $H_i \la 10^{14} GeV$ and
$H_0 \sim 10^{-33} eV$ \cite{tomwilly}. Varying the parameter
$\kappa$ one finds qualitatively similar bounds.  When the
inequality (\ref{bound}) is violated, \cite{tomwilly} conclude
that the universe would eventually end in a big crunch, despite
the fact that standard FRW evolution would indicate otherwise.

\begin{figure}[thb!]
\hspace{5.2truecm}
\epsfysize=2.4truein \epsfbox{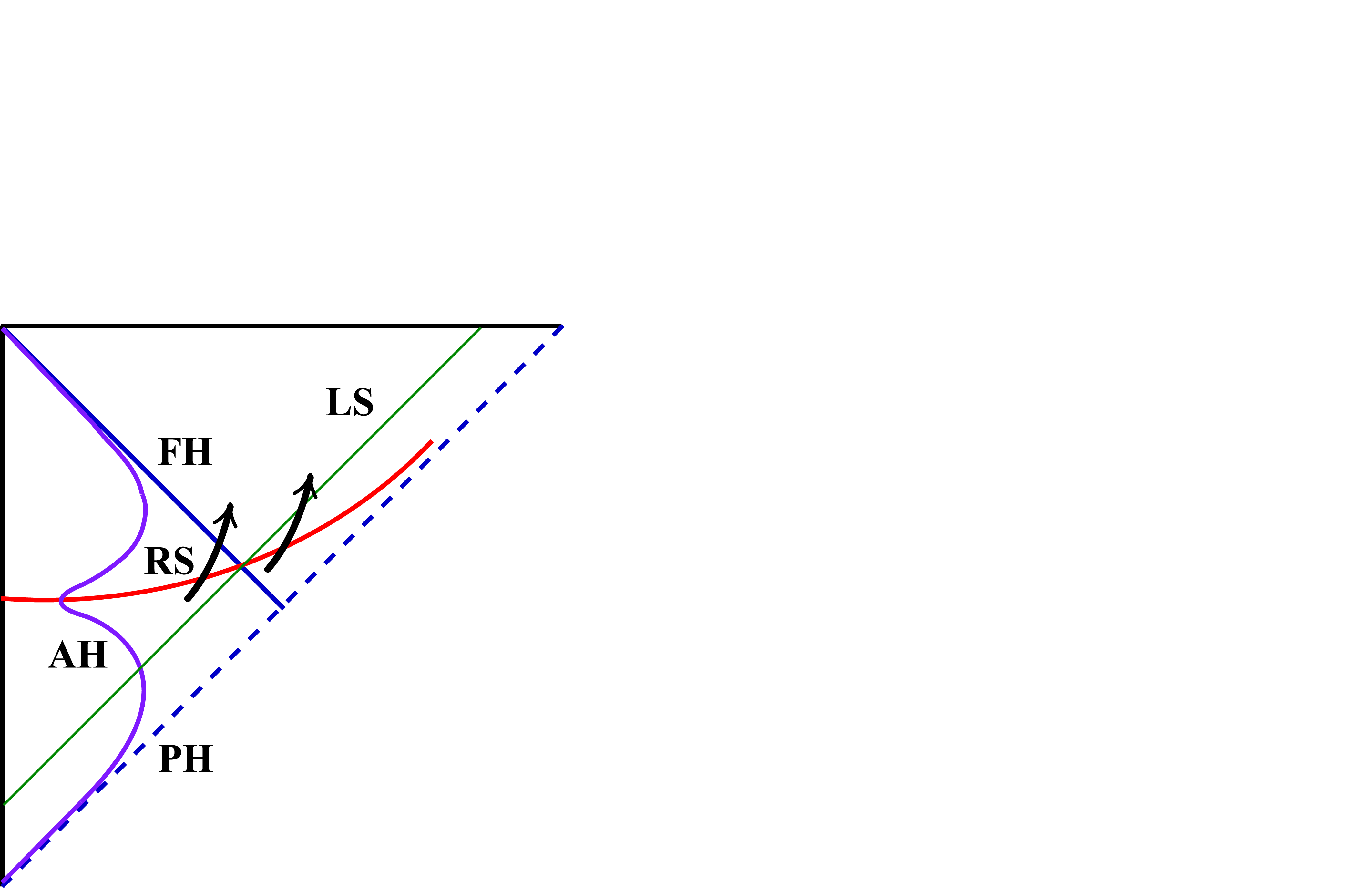}
\caption[]%
{\small\sl Causal patch of an observer in a universe where
inflation and reheating are followed by eternal accelerated
expansion. The symbols designate: PH and FH - past and future
sections of the event horizon, AH - apparent, or Hubble, horizon,
RS - reheating surface, and LS - a future oriented light sheet,
which intersects both the event horizon and the infinitely
inflated future. Black arrows are the worldlines of the entropy
released at the end of inflation.  } \label{f1}
\end{figure}
The situation is shown in Fig \ref{f1}., where we depict the
conformal spacetime diagram for a universe which undergoes early
inflation, reheats and goes through the era of radiation and
matter domination, and finally accelerates forever. Note that the
entropy generated by reheating is uniformly deposited along the
reheating surface RS, which corresponds to a spatially flat slice
at the moment when inflation ends, at least in a region of size $
\sim e^{N}/H_i$. The size of this region can greatly exceed the
Hubble scale $1/H_0$ during the late acceleration era if $N$ is
large. After a long inflation, an initial Hubble patch will expand
to fill a volume much larger than the volume inside the event
horizon, and therefore the reheating surface will extend well past
the future Hubble horizon $1/H_0$ and the event horizon (see Fig.
\ref{f1}.). Computing the entropy as in equation (\ref{entropy})
corresponds to counting all the entropy along the reheating
surface, including the parts both inside and outside the causal
patch. Thus if inflation is long the entropy (\ref{entropy}) may
exceed greatly the entropy that an observer in an eternally
accelerating universe will ever see. Banks and Fischler argue
\cite{tomwilly} that this implies an incompatibility between the
assumptions of the conventional QFT picture of inflation and the
bound on the number of states in the Hilbert space of an
asymptotically de Sitter universe. They claim this means that the
predictions about observable consequences of models with a large
number of e-folds cannot be trusted.

Here we take the point of view that in backgrounds with horizons,
the entropy contents of the spacetime conform to the covariant
entropy bounds of \cite{willylenny,bousso}. With this assumption,
the existence of a horizon constrains only the entropy {\it
inside} the final Hubble  volume to be less than the area of its
horizon.  It does not restrict the entropy deposited {\em outside}
\footnote{We agree, however, that it is unclear how to think about
the entropy deposited in this large volume from the viewpoint of
any given observer.} of it. This bound is clearly not violated in
our universe, as the entropy of the observable universe is many
orders of magnitude below the horizon area.  To an observer inside
the causal patch in Fig. \ref{f1}., the portion of the reheating
surface RS which is outside the event horizon lies beyond her
causal future, in a sense. She will never see that inflation ended
there and that any entropy has been released, or for that matter,
that inflation even happened there.

Outside of the event horizon, then, we should constrain the
entropy on the part of the reheating surface there using the
lightsheet labelled LS in Fig. \ref{f1}. According to the
covariant bound \cite{bousso}, the entropy that crosses any
segment of a lightsheet is bounded by a maximum of the area along
this segment, in Planck units. However, because LS intersects the
infinitely inflated future at the top of the diagram, and so its
maximal area diverges, the covariant entropy bound does not give
an interesting constraint. It can easily accommodate the entropy
released after arbitrarily long inflation. For a more detailed
discussion of these issues, see \cite{marolflowe}.
Problems with adding up entropy stored on large and
arbitrarily chosen spacelike surfaces have been noticed before
\cite{willylenny,bousso,spatials}, and they are consistently
resolved by the application of the covariant bound.

\section{How Many E-folds Can We See?}

\begin{figure}[thb!]
\hspace{2.2truecm}
\epsfysize=2.4truein \epsfbox{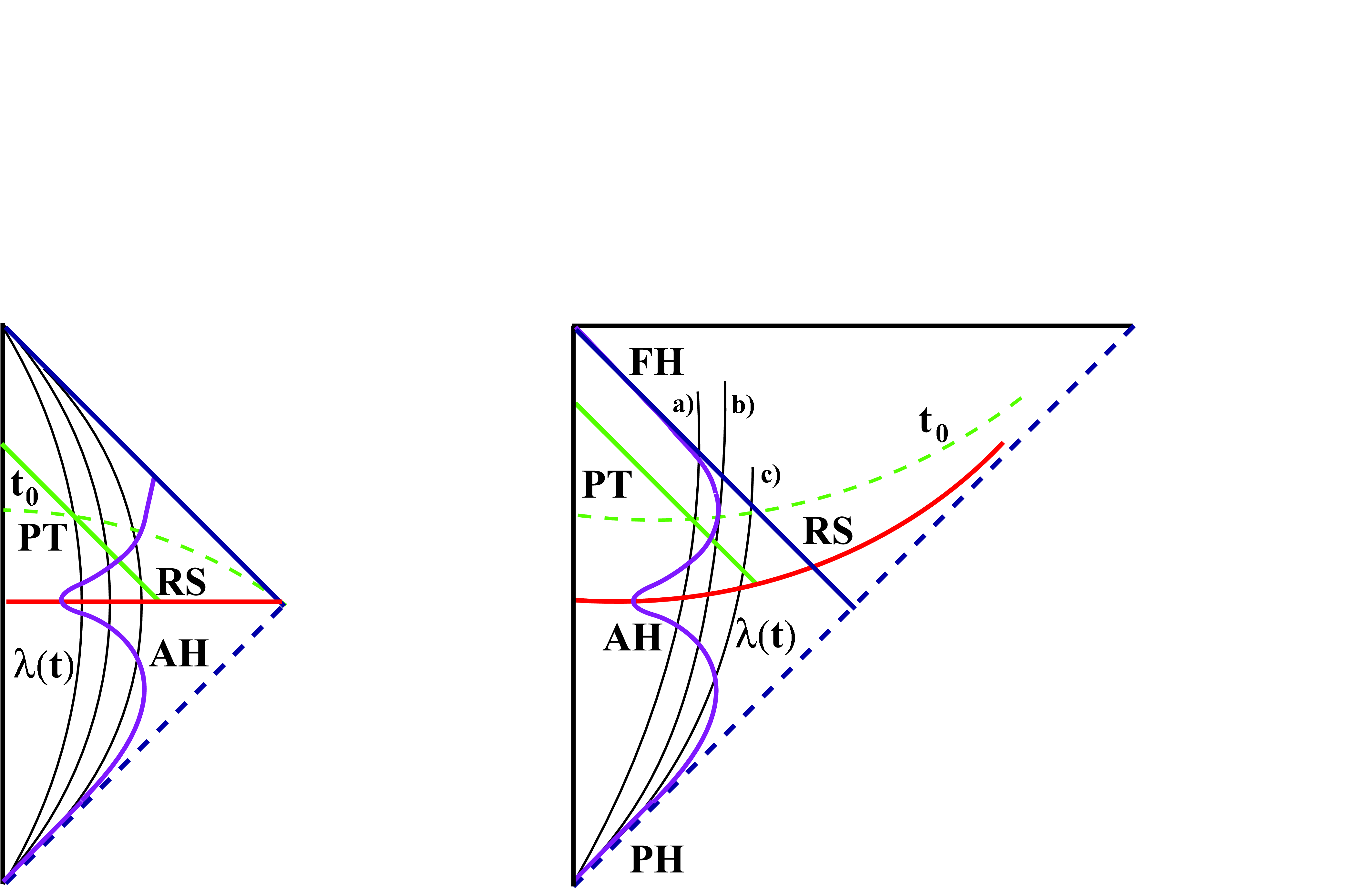}
\caption[]%
{\small\sl Evolution of the wavelengths of some typical
inflationary perturbations in the causal patch in a universe
without (left panel) and with (right panel) event horizons. In the
left panel, all fluctuations eventually reenter the Hubble
horizon. In the right panel, in the case $a)$, a fluctuation is
stretched outside of the Hubble horizon during inflation, remains
there for a time, then reenters during a matter dominated era
after inflation, and eventually gets expelled out of the horizon
once more during the final stage of acceleration. In the case b),
the fluctuation could have reentered about now, but the late
acceleration pushes it back out. In the case c), the late
acceleration prevents the fluctuation from ever reentering the
Hubble horizon.} \label{wave}
\end{figure}

In this section we argue that if in the future the spacetime
undergoes eternal acceleration, the event horizon limits the total
number of e-folds that we can ever {\em observe} to the last $N
\sim 60$ or so. Recall that we study inflation by observing
temperature and density contrasts on the sky. The contrasts at
larger scales correspond to fluctuations that were produced
earlier in inflation. In order to solve the horizon and flatness
problems, inflation must have lasted at least $N \sim 60$ e-folds
\cite{inflation}. Quantum fluctuations during this stage are
imprinted on the curvature via a mechanism closely related to
gravitational particle production, and are subsequently stretched
by inflation to super (Hubble) horizon scales \cite{bard,perts}.
Once there, they ``freeze out", i.e. their amplitude approaches a
constant set by the horizon crossing condition, and their
wavelength scales with the particle horizon, $\lambda(t) =
\lambda_0 \, a(t)/a_0$. What happens to them next depends on the
subsequent evolution of the universe. If inflation ends and
reheating occurs, the Hubble horizon starts growing linearly in
time, while the wavelength stretches more slowly, as $\lambda \sim
a(t)$. If the vacuum energy is zero, this situation will persist
indefinitely, and after a long enough time the Hubble horizon
catches up with the perturbation (see the left panel of Fig.
\ref{wave}), after which the perturbation ``melts"; i.e., it
begins to oscillate and to seed structure formation via the
Jeans instability \cite{bard,perts}.
A patient observer in such a universe would be able to see
arbitrarily far back into inflation: the longer she waits, the
earlier the fluctuations she sees were created.

However, if at some time the post-inflationary universe begins to
accelerate and continues to do so forever, there will be event
horizon as in the right panel of Fig. \ref{wave}. In this case a
(huge!) part of the global spacetime is permanently inaccessible
to any given observer. The evolution of inflationary perturbations
is very different in this case. Depending on when they are
produced, inflationary fluctuations could either (see the right
panel of Fig. \ref{wave}): a) reenter the Hubble horizon during
matter domination, and then eventually be expelled again in the
future, b) in the marginal case, have a wavelength which equals
the Hubble horizon size at about the time when the universe begins
to accelerate again, or c) never reenter, and remain outside the
Hubble horizon forever after their eviction from it during early
inflation.

Inflationary fluctuations that reenter the horizon produce small
curvature perturbations on the background geometry, generating a
distribution of gravitational potential wells. As a result, an
observer can gain information about inflation by examining the
structures which form by the accretion of matter in these
gravitational wells, and by observing anisotropies induced by
these wells in the temperature of the thermal photons released
during reheating.\footnote{For simplicity we ignore the difference
between the reheating surface and the last scattering surface
here.} As time goes on, an observer in an accelerating universe
will notice a gradual loss of the information about inflation. She
will observe a lack of new structures at the largest scales,
because the inflationary fluctuations stop reentering after the
onset of late acceleration. Further, she will notice that the
structures that have already begun forming start to disperse, as
fluctuations at sub-horizon scales get stretched out to larger and
larger distances. Eventually all the inflationary fluctuations
which re-entered during radiation and matter domination will be
pushed out of the Hubble horizon, whose interior will be smoothed
out again (at least on large scales).

However, the photons which comprise the CMB originate on the slice
(i.e. a sphere) of the last scattering surface (RS) which is
separated from the observer by null geodesics (labelled PT, for
photon trajectory, in Fig. \ref{wave}.). The inflationary
fluctuations are imprinted on them en route to the observer via
the Sachs-Wolfe effect, and appear as a distribution of hot and
cold spots on the last scattering surface. In a decelerating
universe, the radius of this last scattering sphere grows without
bound, the pattern of spots changes, and new information about
inflation continues to become available over time. Eventually, if
one continued to observe the pattern of anisotropies in the CMB,
the entire history of the inflationary period would (in principle)
be available.

In a universe which accelerates, the last-scattering sphere
asymptotes to the size of the event horizon at the time of last
scattering, which is finite.  As the acceleration continues,
waiting a given period of time will correspond to a smaller and
smaller change in the size of the last-scattering sphere.
Therefore, the pattern of anisotropies in the CMB will ``freeze"
after the transition to future acceleration, first on the largest
scales, and then on shorter and shorter scales. Continuing to
observe after the beginning of the late acceleration will not
reveal any information about periods of inflation earlier than
those that have already been seen, and will at best slightly
improve the data on the already visible period.\footnote{We would
like to thank Gil Holder for discussions on this point.} The
acceleration freezes an ever-fainter image of one slice of the
last scattering surface on the sky, for a very long time.

Eventually, however, even this information will be erased.
Spacetimes with event horizons contain Hawking particles, and as
the cosmological expansion advances, the CMB cools until it
reaches a point where the number of CMB photons counted by an
observer drops below the number of Hawking photons. After this
time, any information in the CMB will be masked by the ``noise" in
this cosmological Hawking radiation. Asymptotically the bath of
Hawking particles will completely overwhelm the CMB. Some
implications of the loss of a record of the last stages of
inflation for astronomy have been discussed in
\cite{gudbjo,loeb,Busha}.

Let us now quantify our bound. First, we recall the derivation of
the minimum number of e-folds necessary to solve the horizon and
flatness problems \cite{inflation}.  Let us again assume that we
begin with one Hubble patch of homogeneous space. Inflation must
then produce a sufficiently large number $N$ of e-folds such that
this initial patch evolves into a region the size of the present
Hubble horizon size, $(H_0)^{-1} \sim (10^{-33} eV)^{-1} $. The
wavelengths of perturbations grow in time according to
\be \lambda(t) = \lambda_0 \, \frac{a(t)}{a(t_0)} \, .
\label{wavelength}\ee
Taking $t_0$ to be ${\cal O}$(today) ($t_0 \sim 10^{10}$ years),
we are interested in the largest scale observable now,
namely\footnote{We denote quantities evaluated at time $t = t_0$
with a subscript $0$.} $\lambda_0 = 1/H_0$. A horizon scale
perturbation originated during inflation at some time $t_b < t_0$,
when its wavelength was the inflationary Hubble size,
$\lambda(t_b) = 1 / H(t_b)$.  Hence,
\be a(t_b) H(t_b) = a_0 H_0 \, . \label{matching} \ee
Equation (\ref{matching}) is the usual horizon crossing condition
\cite{perts} in a slightly unconventional form. Approximating the
inflating phase as de Sitter space with a constant Hubble scale
$H_i$ and using the flat slicing, we have $a(t) = a_e \exp(H_i(t -
t_e))$ for times during inflation, where the subscript $e$ refers
to the end of inflation. Evaluating this at a time $t_b$ during
inflation and substituting into (\ref{matching}) yields
\be N \equiv H_i (t_e - t_b) = \ln \left( {a_e H_i \over a_0 H_0}
\right) \, . \label{efolds} \ee
After inflation, the universe grew by a factor of about $a_0/a_e
\sim T_e/T_0$, where $T_e$ is the reheating temperature and $T_0
\sim 10^{-3} eV$ the current CMB temperature. Taking this ratio to
be about $10^{26}-10^{28}$ and the scale of inflation to be $H_i
\la 10^{14} GeV$, one finds $N \sim 60$, with some sensitivity on
the reheating temperature, the scale of inflation et cetera, which
we will ignore here (see \cite{inflation}).

\begin{figure}[thb!]
\hspace{-0.2truecm}
\epsfysize=2.2truein \epsfbox{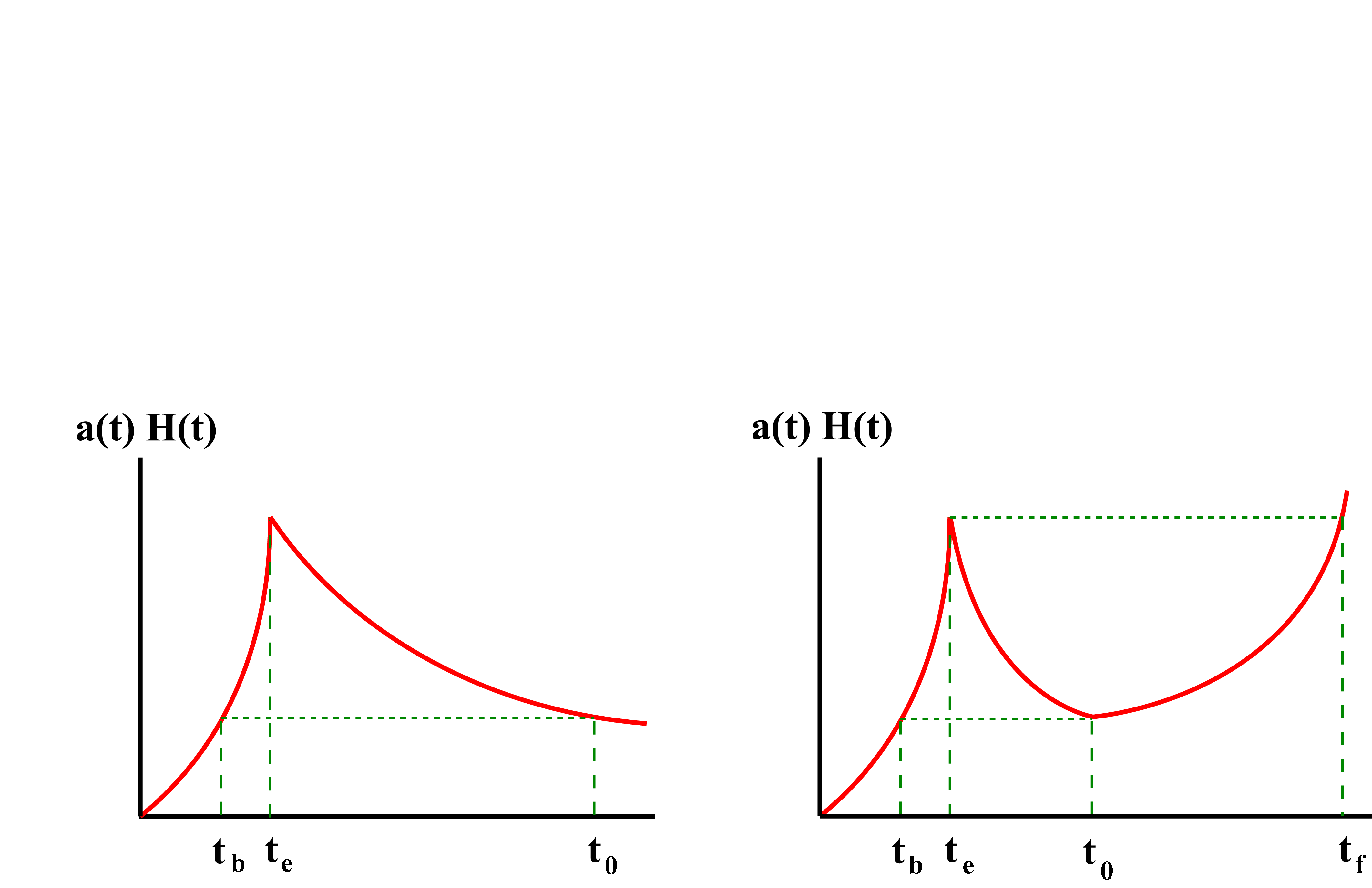}
\caption[]%
{\small\sl On the left, evolution of the comoving Hubble scale
$a(t) H(t)$ for a universe which inflates, followed by radiation
and matter domination; on the right, the same graph for a universe
that enters a late-time accelerating phase. } \label{scale3}
\end{figure}
To make this equation more transparent, we plot the comoving
Hubble scale $a(t) H(t)$ for a universe without any late epoch of
cosmic acceleration in the left panel of Fig. \ref{scale3}.
Initially it grows exponentially because of inflation.
Subsequently, it decreases as a small negative power of $t$,
because after reheating the universe decelerates. For example, if
the universe is dominated by matter with an equation of state $p =
w \rho$, $a(t) H(t)$ scales as $t^{-(1 +3w)/[3(1 + w)]}$, which is
decreasing for $w > -1/3$. As $a(t) H(t)$ decreases, it scans
through more and more values of the comoving momentum $k =
1/\lambda_0$, which means that those scales reenter the horizon.
Thus, regardless of how large a scale $\lambda_0$ is, if the
universe decelerates forever and $a(t) H(t)$ continues to
decrease, at some time this scale will reenter the Hubble horizon.

On the other hand, if the universe accelerates in the future, the
comoving Hubble scale $a(t) H(t)$ begins to grow again at late
times, as we can see by setting $w < -1/3$ in the scaling law
given above. At a time $t_f$, where $f$ stands for final, when the
comoving Hubble scale equals its value at reheating, the very last
perturbation generated during inflation will be pushed back out of
the horizon. Indeed, after $t_f$ no inflationary perturbations
will remain in the Hubble horizon and no new structure will form
from the seeds generated by inflation\footnote{In reality, the
information about the primordial inflation encoded in the shortest
scales generated during inflation will already be strongly
contaminated by the nonlinear effects occurring in the intervening
period between $t_e$ and $t_f$, such as galaxies, clusters etc. We
are ignoring this contamination here.} (see the right panel of
Fig. \ref{scale3}.). The time $t_f$ is defined by the equality
\be a(t_f) H(t_f) = a(t_e) H_i \, , \label{unobservable} \ee
where $t_e$ is the time at reheating.  The value of $t_f$ depends
on the equation of state of the dark energy in a way we calculate
below.

As we have mentioned above, spacetimes with event horizons contain
Hawking particles, with a characteristic temperature\footnote{This
is certainly correct for a positive cosmological constant ($w =
-1$). For quintessential universes, we believe there is a similar
effect \cite{hks}, but we are not aware of a precise calculation
of it.} given by the Gibbons-Hawking formula $T_H = H/ 2\pi$
\cite{gehaw}. This temperature does not redshift in the usual way,
because the Hawking radiation is continuously replenished by
quantum fluctuations, rather than being a remnant of an earlier
hot big bang. If the universe is accelerating, the CMB temperature
$T_{CMB}$ will eventually redshift to a point where it is equal to
$T_H \sim H(t)$.  If $T_e$ was the reheating temperature, under
adiabatic evolution the temperature at any later time is related
to it by $T_{CMB}(t) = T_e a(t_e)/a(t)$. Thus the temperatures of
the CMB and the Hawking particles obey $T_{CMB}(t_T) = T_H$ at a
time $t_T$ when
\be  a(t_T) H(t_T) = a(t_e) T_e \label{reheat} \, . \ee
If reheating were perfectly efficient, the reheating temperature
would be related to the Hubble scale at the end of inflation by
$T_e \sim \sqrt{H_i}$ (recall that we have set the Planck mass
equal to unity). Thus, since $H_i < 1$, $T_e > H_i$, and $t_T >
t_f$. In practice, however, the reheating temperature is model
dependent\footnote{As a result, the right-hand side of the
equation (\ref{reheat}) should really read $a(t_R) T_R$, and this
quantity may evolve slightly differently. However, the differences
will all be model dependent and confined to short scales, and so
we will ignore this here.}. It is possible that there are some
models where the ordering of $t_f$ and $t_T$ is reversed. But in
this case, for $t> t_T$ the CMB photons will be outnumbered by
Hawking photons, and it would be impossible to extract any
information about inflation from their fluctuations.

Therefore if the cosmic acceleration never ends, only those
inflationary fluctuations with comoving momenta in the interval
$a(t_b) H(t_b) \le k \le a(t_e) H(t_e)$ will ever be observable.
An observer will never be able to see much past the last 60
e-folds of inflation, however patient she may be. Further, the
information which was accessible to her will be lost after the
time $t_T$, the value of which depends on the equation of state of
the dark energy ($t_T \sim t_f$). It can be found by solving
(\ref{reheat}),
eqs. (\ref{matching}) and (\ref{efolds}), and the scaling
$a(t) H(t) \sim a_0 H_0 (t/t_0)^{-(1 +3w)/[3(1 + w)]}$ when $-1<w<-1/3$:
\be \label{finaltime} t_T \sim 10^{78 (1+w)/|1+3w|} t_0 \, . \ee
In the limit $w \rightarrow -1/3$ the time diverges, as expected
since for $w \ge -1/3$ the event horizon and the Hawking particles
disappear, and the information about early inflation survives and
remains available to an investigation by a patient observer. The
limit $w \rightarrow -1$ is simpler to determine by directly
substituting $a(t_T) H(t_T) = a_0 H_0 \exp(H_0 t_T)$, which yields
\be t_T \sim \frac{60}{H_0} \, . \label{cc} \ee
Hence if the dark energy is a small cosmological constant, the
record of early inflation will be lost in about a trillion years
(see also \cite{gudbjo,loeb,Busha} for astronomical implications
of these time scales).

\section{Summary}

In closing, we note that because of the Grishchuk-Zel'dovich
effect \cite{grize} the perturbations at superhorizon scales may
have a weak effect on observable structures. The absence of large
perturbations at the current horizon scale implies that the
momentum space scale below which the perturbation power spectrum
may change significantly must be about 500 times lower than the
current Hubble scale $H_0$ \cite{turner,Grishchuk,Kashlinsky}. A
more recent analysis of the WMAP data improves this limit to be
about $4000$ times lower than the Hubble scale \cite{Castro}. This
allows us to probe the period of inflation slightly more than 60
e-folds from the end, giving us some information about perhaps 8
more.

To conclude, having assumed the covariant entropy bounds
\cite{willylenny,bousso} we have found no limitations on the
number of allowed e-folds of inflation in universes dominated at
late times by dark energy. However, we have found that eternal
dark energy with $w < -1/3$ does prevent us from ever measuring
inflationary perturbations which originated before the ones
currently observable. Further, it slowly degrades the information
stored in the currently observable perturbations. This allows us
to re-formulate the ``Why Now?" problem in a novel and interesting
way: why are we living in the time at which we can see back to the
earliest scales?  In other words, why would the number of e-folds
required to solve the horizon problem and explain the observed
large scale homogeneity, isotropy and flatness of the universe
also be the maximum number of e-folds which we will ever be able
to observe?

\vspace{1.5cm}

{\bf \noindent Acknowledgements}

\smallskip

We thank A. Albrecht, T. Banks, W. Fischler, G. Holder, M.
Kaplinghat, A. Linde, M. Sloth, L. Susskind, S. Thomas, and T.
Tyson for useful discussions. The work of NK and LS was supported
in part by the DOE Grant DE-FG03-91ER40674, in part by the NSF
Grant PHY-0332258 and in part by a Research Innovation Award from
the Research Corporation. The work of MK is supported by NSF grant
PHY-0070928. MK thanks the UC Davis cosmology group for
hospitality during this work.


\begin{thebibliography}{99}

\bibitem{sne}
A.~G.~Riess {\it et al.},
Astron.\ J.\  {\bf 116} (1998) 1009;
S.~Perlmutter {\it et al.},
Astrophys.\ J.\  {\bf 517} (1999) 565.

\bibitem{new}
J.~L.~Tonry {\it et al.},
Astrophys.\ J.\  {\bf 594}, (2003) 1;
R.~A.~Knop {\it et al.},
astro-ph/0309368;
A.~G.~Riess {\it et al.},
astro-ph/0402512.

\bibitem{cmb}
A.~H.~Jaffe {\it et al.},
Phys.\ Rev.\ Lett.\  {\bf 86} (2001) 3475;
A.~E.~Lange {\it et al.},
Phys.\ Rev.\ D {\bf 63} (2001) 042001;
A.~Balbi {\it et al.},
Astrophys.\ J.\  {\bf 545} (2000) L1;
D.~N.~Spergel {\it et al.},
astro-ph/0302209.

\bibitem{weinberg}
S.~Weinberg,
Rev.\ Mod.\ Phys.\  {\bf 61} (1989) 1.

\bibitem{q} C.~Wetterich,
Nucl.\ Phys.\ B {\bf 302} (1988) 668;
B.~Ratra and P.~J.~E. Peebles, Phys.\ Rev.\ D {\bf 37} (1988)
3406;
R.~R.~Caldwell {\it et al.}, Phys.\ Rev.\ Lett.\ {\bf 80} (1988)
1582;
L.~Wang {\it et al.}, Astrophys.\ J.\ {\bf 530} (2000) 17;
A.~Albrecht and C.~Skordis,
Phys.\ Rev.\ Lett.\  {\bf 84} (2000) 2076;
C.~Armendariz-Picon, V.~Mukhanov and P.~J.~Steinhardt,
Phys.\ Rev.\ Lett.\  {\bf 85} (2000) 4438.

\bibitem{tom}
T.~Banks,
hep-th/0007146;
hep-th/0011255;
T.~Banks and W.~Fischler,
hep-th/0102077.

\bibitem{hks}
S. Hellerman, N. Kaloper and L. Susskind,
JHEP {\bf 0106} (2001) 003; W. Fischler, A. Kashani-Poor, R.
McNees and S. Paban,
JHEP {\bf 0107} (2001) 003.

\bibitem{witten}
E.~Witten,
hep-ph/0002297.

\bibitem{cosconst}
L.~Dyson, M.~Kleban and L.~Susskind,
JHEP {\bf 0210} (2002) 011;
T.~Banks, W.~Fischler and S.~Paban,
JHEP {\bf 0212} (2002) 062;
N.~Goheer, M.~Kleban and L.~Susskind,
JHEP {\bf 0307} (2003) 056.

\bibitem{tomwilly}
T.~Banks and W.~Fischler,
astro-ph/0307459.

\bibitem{marolflowe}
D.~A.~Lowe and D.~Marolf,
hep-th/0402162.

\bibitem{willylenny}
W.~Fischler and L.~Susskind,
hep-th/9806039.

\bibitem{bousso}
R.~Bousso,
JHEP {\bf 9907} (1999) 004;
JHEP {\bf 9906} (1999) 028;
JHEP {\bf 0011} (2000) 038.

\bibitem{spatials}
D.~Bak and S.~J.~Rey,
Class.\ Quant.\ Grav.\  {\bf 17} (2000) L83;
R.~Easther and D.~A.~Lowe,
Phys.\ Rev.\ Lett.\  {\bf 82} (1999) 4967;
G.~Veneziano,
Phys.\ Lett.\ B {\bf 454} (1999) 22;
N.~Kaloper and A.~D.~Linde,
Phys.\ Rev.\ D {\bf 60} (1999) 103509;
R.~Brustein and G.~Veneziano,
Phys.\ Rev.\ Lett.\  {\bf 84} (2000) 5695.

\bibitem{inflation}
A. Guth,
Phys.\ Rev.\ D {\bf 23} (1981) 347;
A. Linde,
Phys.\ Lett.\ B {\bf 108} (1982) 389;
A. Albrecht and P. Steinhardt,
Phys.\ Rev.\ Lett.\  {\bf 48} (1982) 1220.

\bibitem{bard}
J.~M.~Bardeen,
Phys.\ Rev.\ D {\bf 22} (1980) 1882.

\bibitem{perts}
V. Mukhanov and G. Chibisov,
JETP Lett.\  {\bf 33} (1981) 532;
A. Starobinsky,
Phys.\ Lett.\ B {\bf 117} (1982) 175;
S. Hawking,
Phys.\ Lett.\ B {\bf 115} (1982) 295;
A. Guth and S.-Y. Pi,
Phys.\ Rev.\ Lett.\  {\bf 49} (1982) 1110;
J. Bardeen, P. Steinhardt and M. Turner,
Phys.\ Rev.\ D {\bf 28} (1983) 679;
A. Linde,
Phys.\ Lett.\ B {\bf 116} (1982) 335;
M.~Sasaki,
Prog.\ Theor.\ Phys.\  {\bf 70}
(1983) 394;
V. Mukhanov,
JETP Lett.\  {\bf 41} (1985) 493.

\bibitem{gudbjo}
E.~H.~Gudmundsson and G.~Bjornsson,
Astrophys.\ J.\  {\bf 565} (2002) 1.


\bibitem{loeb}
A.~Loeb,
Phys.\ Rev.\ D {\bf 65} (2002) 047301;
K.~Nagamine and A.~Loeb,
New Astron.\  {\bf 8} (2003) 439;
astro-ph/0310505.

\bibitem{Busha}
M.~T.~Busha, F.~C.~Adams, R.~H.~Wechsler and A.~E.~Evrard,
Astrophys.\ J.\  {\bf 596} (2003) 713;
F.~C.~Adams, M.~T.~Busha, A.~E.~Evrard and R.~H.~Wechsler,
Int.\ J.\ Mod.\ Phys.\ D {\bf 12} (2003) 1743.

\bibitem{gehaw}
G.~W.~Gibbons and S.~W.~Hawking,
Phys.\ Rev.\ D {\bf 15} (1977) 2738.

\bibitem{grize}
L. Grischuk and Ya. B. Zel'dovich, Sov. Astron. {\bf 22} (1978)
125.

\bibitem{turner}
M.~S.~Turner,
Phys.\ Rev.\ D {\bf 44} (1991) 3737.


\bibitem{Grishchuk}
L.~P.~Grishchuk,
Phys.\ Rev.\ D {\bf 45} (1992) 4717.


\bibitem{Kashlinsky}
A.~Kashlinsky, I.~I.~Tkachev and J.~Frieman,
Phys.\ Rev.\ Lett.\  {\bf 73} (1994) 1582.

\bibitem{Castro}
P.~G.~Castro, M.~Douspis and P.~G.~Ferreira,
Phys.\ Rev.\ D {\bf 68} (2003) 127301.


\end{thebibliography}
\end{document}